\documentclass[pra,aps,twocolumn,amsmath,amssymb]{revtex4}
\usepackage{graphicx}
\pdfoutput=1
\usepackage{amsmath}
\usepackage{color}
\usepackage{float}
\setcitestyle{super}
\usepackage[english]{babel}
\usepackage{color,soul}

\begin{document}
\title{Conventional Superconductivity in Type II Dirac Semimetal PdTe$_2$ }

\author{Shekhar Das, Amit, Anshu Sirohi, Lalit Yadav, Sirshendu Gayen}

\author{Yogesh Singh}
\email[Contact for materials, E-mail: ]{yogesh@iisermohali.ac.in}

\author{Goutam Sheet}
\email[Corresponding author, E-mail: ]{goutam@iisermohali.ac.in}

\affiliation{Department of Physical Sciences, Indian Institute of Science Education and Research(IISER), Mohali, Sector 81, S. A. S. Nagar, Manauli, PO: 140306, India.}

\begin{abstract}

\textbf{The transition metal dichalcogenide PdTe$_2$ was recently shown to be a unique system where a type II Dirac semimetallic phase and a superconducting phase co-exist. This observation has led to wide speculation on the possibility of the emergence of an unconventional topological superconducting phase in PdTe$_2$. Here, through direct measurement of the superconducting energy gap by scanning tunneling spectroscopy (STS), and temperature and magnetic field evolution of the same, we show that the superconducting phase in PdTe$_2$ is conventional in nature. The superconducting energy gap is measured to be 326 $\mu$eV at 0.38 K and it follows a temperature dependence that is well described within the framework of Bardeen-Cooper-Schriefer's (BCS) theory of conventional superconductivity. This is surprising because our quantum oscillation measurements confirm that at least one of the bands participating in transport has topologically non-trivial character.}

\end{abstract}

\maketitle

It is believed that a superconducting ground state realized on a material displaying topologically non-trivial character might lead to the discovery of the elusive phase of matter called topological superconductivity\cite{Ludwig, Zhang1, Zhang2} where relativistic Majorana fermions could emerge as excitations. \cite{Kane, YAndo, SDS1, SDS2, SDS3, SDS4, Leo1} In the past, attempts were made to realize such a phase of matter through a number of ways, like, doping a topological insulator,\cite{CuBi2Se3, Smylie1, Smylie2} applying pressure on topological systems,\cite{Bi2Te3} fabricating proximity-coupled heterostructures of topological insulators and conventional superconductors,\cite{Leo1, Leo2} making mesoscopically confined point contacts on topological crystalline insulator,\cite{Sdas} Dirac\cite{NM,Jwei} or Weyl semimetals \cite{NC} etc. In such systems, while a superconducting phase could be achieved, clear manifestation of topological superconductivity remained an unattained goal. Most recently, with the discovery of a type-II Dirac semimetallic phase on the dichalcogenide PdTe$_2$,\cite{Park, Fei, Duan} where a superconducting phase is known to emerge naturally,\cite{Finlayson, Roberts} there have been wide speculation that this superconducting phase might eventually emerge as a topological superconductor. Owing to the existence of a topological phase along with superconductivity, it warrants a thorough investigation of the superconducting phase of PdTe$_2$ through high-resolution spectroscopic tools. 

In this Letter, we report scanning tunneling microscopy and spectroscopy results performed on high quality single crystals of PdTe$_2$ in ultra-high vacuum and sub-kelvin temperatures to spectroscopically probe the nature of superconductivity. Through systematic temperature and magnetic field dependent experiments we found that the superconducting phase in PdTe$_2$, despite it's proximity to a topologically non-trivial Dirac semimetallic phase, is conventional in nature where the superconducting order parameter shows a temperature dependence consistent with the prediction of the theory of Bardeen, Cooper and Schriefer (BCS) for conventional superconductors.\cite{BCS}

High quality single crystals of PdTe$_2$ were grown by a melt growth method. The starting elements, Pd ($99.99$\% purity) and Te ($99.9999$\%), were weighed in the atomic ratio $1:2.2$ and sealed in an evacuated quartz tube.  For crystal growth, the sealed quartz tube was heated to 790$^o$C in 15 h, kept there for 48 h, and then it was slowly cooled to 500$^o$C over 7 days. They were then annealed at 500$^o$C for 5 days before cooling naturally. The shiny crystals of a few millimeter size thus obtained could be cleaved easily from the as grown boule. The chemical composition of crystals was verified by energy dispersive spectroscopy (EDS) on a JEOL Scanning Electron Microscope (SEM). The ratio given by EDS between Pd and Te was $1:1.99$, showing the stoichiometric ratio of the compound. Few crystals were crushed into powder for X-ray diffraction measurements. The powder X-ray diffraction pattern confirmed the phase purity of PdTe$_2$, well crystallized in CdI$_2$-type structure with the P$\overline{3}$m1(164) space group. 

The high quality of the crystals were confirmed by a high residual resistivity ratio (RRR) $\rho_{300K}/\rho_{2K} \sim 75$ (see the $inset$ of Figure 1(a)). This is significantly higher than the ratio ($\sim$10) reported for the PdTe$_2$ crystals used in the past\cite{CSyadav}.  Furthermore, as shown in Figure 1(a), the crystals also exhibited de Haas--van Alphen (dHvA) quantum oscillations at a relatively low magnetic field of 3 Tesla\cite{Fei, dhva} with the magnetic field applied perpendicular to the $c$-axis of the crystal. In Figure 1(b), we show the Landau level index ($n$) plot constructed from the quantum oscillation data. The plot was fitted well by a straight line, the extrapolation to $1/B\sim0$ of which gave a non-zero intercept of 0.42. This indicates that the only band (see the FFT plot in the $inset$ of Figure 1(b)) giving rise to the oscillations for $B\perp c$-axis is topologically non-trivial. It should be noted that since we could record signals only for high values of the Landau level indices, the error in the calculation of the intercept could be relatively high thereby, making the determination of exact the Berry phase difficult. An analysis of the temperature dependence of oscillation amplitude with Lifshitz-Kosevich fitting revealed an effective mass of 0.27$m_0$ for the given band, where $m_0$ is the free electron mass. Frequencies corresponding to other bands were observed when the magnetic field was applied parallel to the $c$-axis of the crystal (see Figure 1(c)). The high quality of the crystals is further confirmed by a sharp anomaly in heat capacity at the superconducting transition (inset of Figure 1(a)).

\begin{figure}[h!]
	\centering
		\includegraphics[width=.4\textwidth]{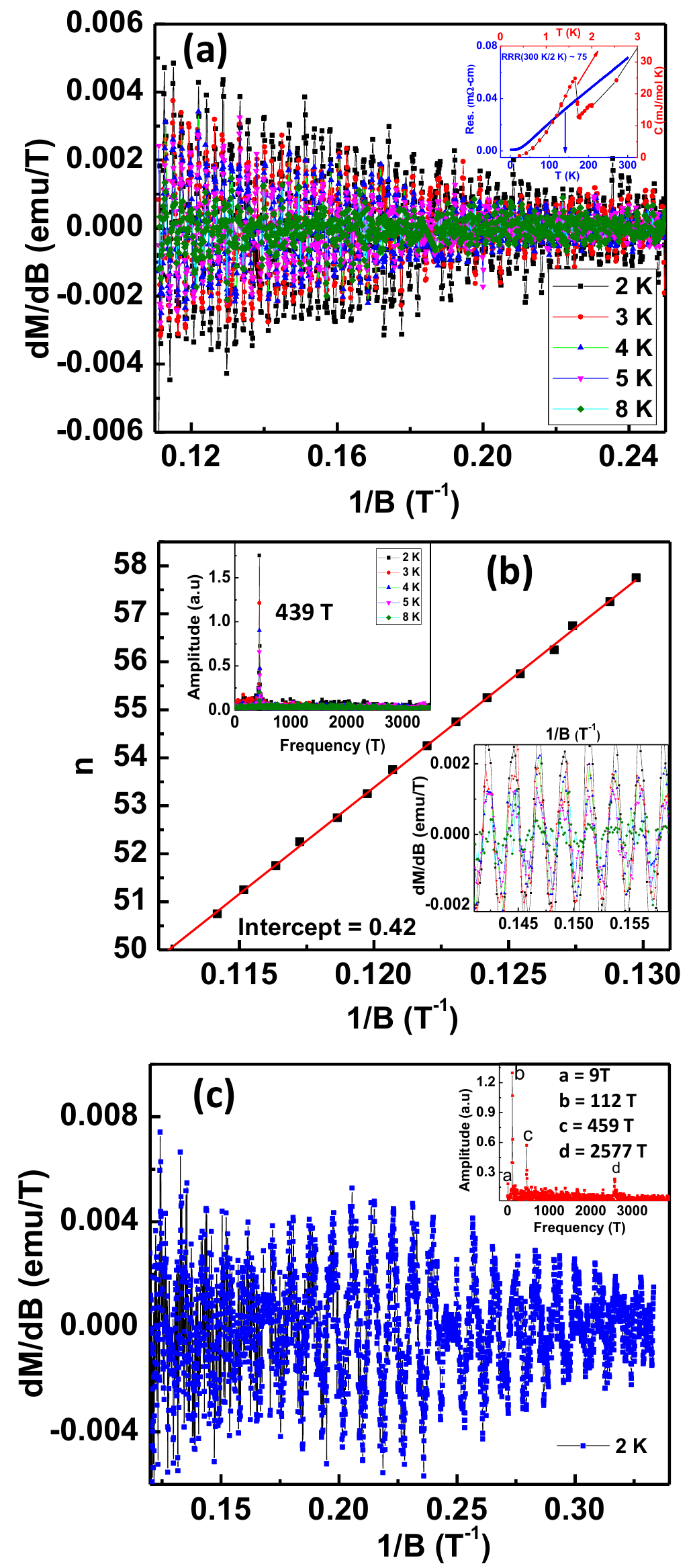}
		\caption{(a)dHvA oscillations at different temperatures measured in a vibrating sample magnetometer (VSM) in a PPMS. Magnetic field was applied perpendicular to the $c$-axis. $inset:$ Resistivity as a function of temperature (left and bottom axes). Heat capacity vs. temperature (top and right axes). (b) Landau level index plot. $insets:$ Upper: FFT of the data shown in (a). The solid line shows a linear fit. Lower: zoomed view of part of the oscillations shown in (a). (c) dHvA oscillations at different temperatures with the field applied along $c$-axis. $inset:$ FFT of the data shown in (c). }
	
	\label{Figure 4}
\end{figure}

The STM and STS experiments were carried out in an ultra-high vacuum (UHV) cryostat working down to 300 mK (Unisoku system with RHK R9 controller). First a single crystal of PdTe$_2$ was mounted in a low-temperature cleaving stage where the crystal was cleaved by an $in-situ$ cleaver at 77 K in UHV ($10^{-11}$mbar). Subsequently, the crystal was transferred by an UHV manipulator to the scanning stage at low-temperature. This process minimized the possibility of contamination and/or modification of the pristine surface. In Figure 2(a,b) we show atomic resolution images of the PdTe$_2$ surface captured at 385 mK. The surface shows a triangular or hcp arrangement of the atoms with an interatomic distance of 4.0$\pm$0.1 \r{A}, which matches well with the previously reported measurements\cite{Ryan}. The inset of Figure 2(b) depicts the topographic modulation corresponding to the line-cut in Figure 2(b). As indicated by white and yellow arrows in Figure 2(a), two kinds of defects are also visible in the images -- one of them being of so-called ``clover-leaf"-shaped defects (shown by white arrow) that are ubiquitously seen on the surfaces of topological insulators like Bi$_2$Se$_3$, Bi$_2$Te$_3$ etc\cite{ZA, Hanaguri}. It should be noted that in earlier STM works on PdTe$_2$, the defect states were not resolved.\cite{STM} Such defects are known to be associated with the adatoms and vacancies on the surface. In PdTe$_2$, the clover-leaf shaped defects could be due to Te-vacancies in the crystals. In topological insulators, the concentration of defects determine the position of the surface Dirac cone in a given crystal with respect to the Fermi energy. The $\frac{dI}{dV}$ vs. $V$ spectra on PdTe$_2$ recorded at 22 K reveal a ``$V$"-shaped feature which might correspond to a Dirac cone and that is seen to appear 110 meV below the Fermi energy. These observations are consistent with the reported type II Dirac semi-metallic nature of PdTe$_2$.\cite{Fei}

\begin{figure}[h!]
	\centering
		\includegraphics[width=.5\textwidth]{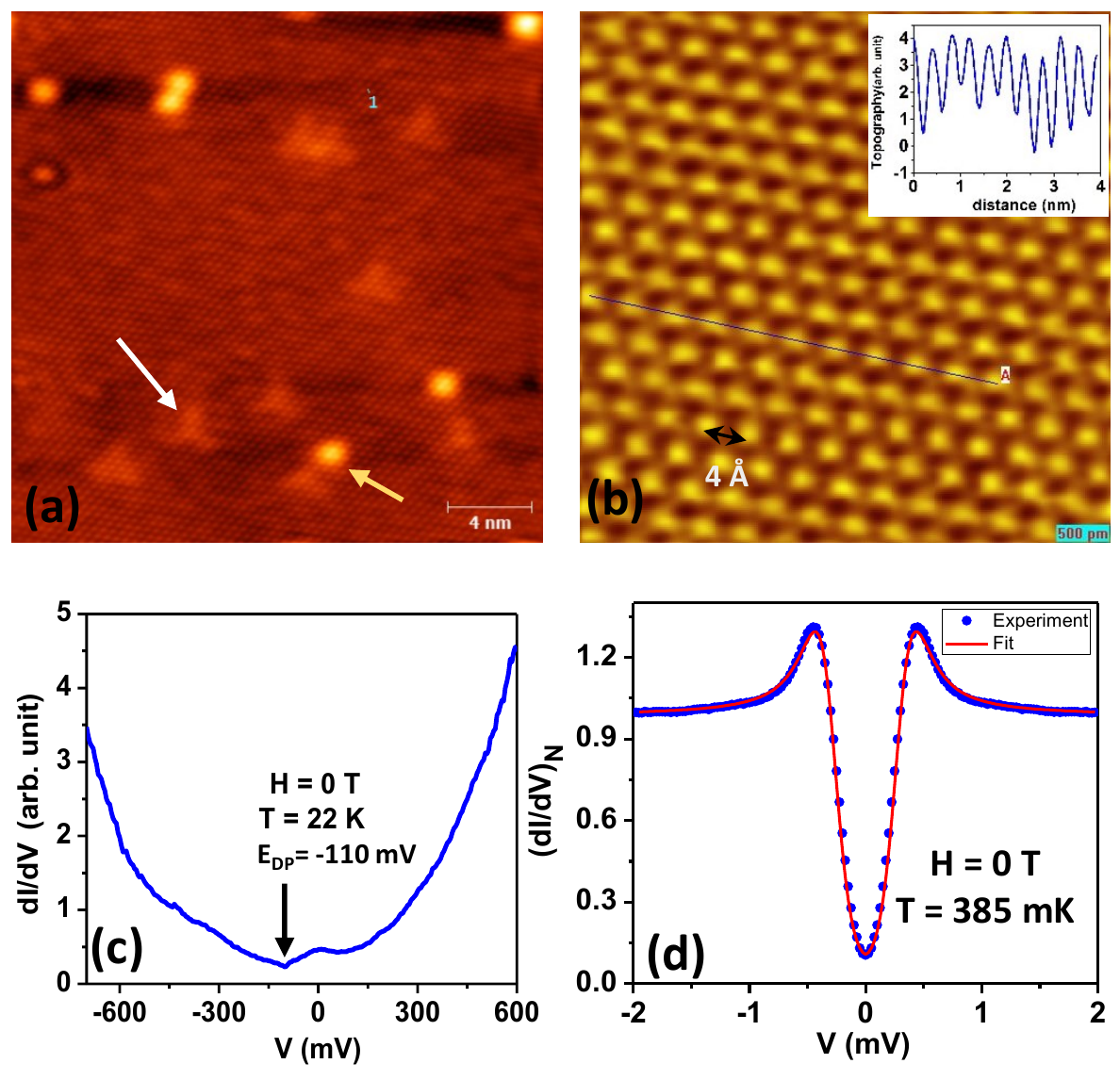}
	\caption{(a) Atomic resolution image of the cleaved surface of PdTe$_2$. The scale bar is 4 nm. Two types of defects are indicated by the white and the yellow arrows respectively. Quasiparticle interference around the defects are also seen. (b) Zoomed image of a defect-free area. The inset shows the topographic modulation along the line-cut in the image. (c) A differential conductance spectrum measured by STS at 22 K. (d) A differential conductance spectrum recorded at 385 mK showing clear coherence peak and a low-bias conductance dip. The open circles represent experimental data and the solid line is a theoretical fit using the tunneling equation for a conventional superconductor with the superconducting energy gap $\left(\Delta \right) = 326$  $\mu eV$.}
	\label{Figure 4}
\end{figure}

After confirming the pristine nature of the surface through atomic resolution imaging, we performed local tunneling spectroscopy at several points on the surface using the STM tip. At lowest temperature, the $\frac{dI}{dV}$ vs. $V$ spectrum shows two clear peaks at $\pm$326 $\mu V$ symmetric about $V=0$, below which $\frac{dI}{dV} \sim 0$ (see Figure 2(d)). These are the coherence peaks due to the superconducting phase of PdTe$_2$ and the peak position provides a direct measure of the superconducting energy gap ($\Delta$). In order to obtain a quantitative estimate of $\Delta$, we fitted the spectrum with the tunneling equation $\frac{dI}{dV} = \frac{d}{dV}\left({G_N}\int_{-\infty}^{+\infty} N_s(E)N_n(E-eV)[f(E) - f(E-eV)]dE\right)$, where $G_N = \frac{dI}{dV}|_{V>>\Delta/e}$, $N_s(E)$ and $N_n(E)$ are the normalized density of states of the BCS-like superconducting sample and the normal metallic tip respectively while $f(E)$ is the Fermi-Dirac distribution function.\cite{Tinkham} Since the relevant energy scale in this case is small ($\sim$ few hundreds of $\mu V$), it is reasonable to assume $N_n(E)$ to be independent of $E$. As per Dyne's formula $N_s(E) = Re\left(\frac{(E-i\Gamma)}{\sqrt{(E-i\Gamma)^2-\Delta^2}}\right)$, where $\Gamma$ is an effective broadening parameter included to take care of slight broadening of the BCS density of states possibly due to finite life time of quasiparticles.\cite{Dynes} In our analysis $\Gamma$ remained very small, within 1\% of $\Delta$. The estimated $\Delta$ at 385 mK is found to be 326 $\mu eV$ thereby giving $\frac{2\Delta}{k_BT_c} \sim$ 2, which falls well within the limit of a weak-coupling conventional BCS superconductor. Therefore, based on the remarkable fitting of the experimental data with the tunneling equation involving BCS-like quasiparticle density of states ($N_s$), we conclude that the observed superconducting phase in PdTe$_2$ is conventional in nature. Furthermore, it should also be noted that apart from the coherence peaks and the smooth decay of the spectrum at higher bias, no other special spectral features are observed. This further indicates the absence of any unconventional component in the superconducting order parameter of PdTe$_2$. 

\begin{figure}[h!]
	\centering
		\includegraphics[width=.365\textwidth]{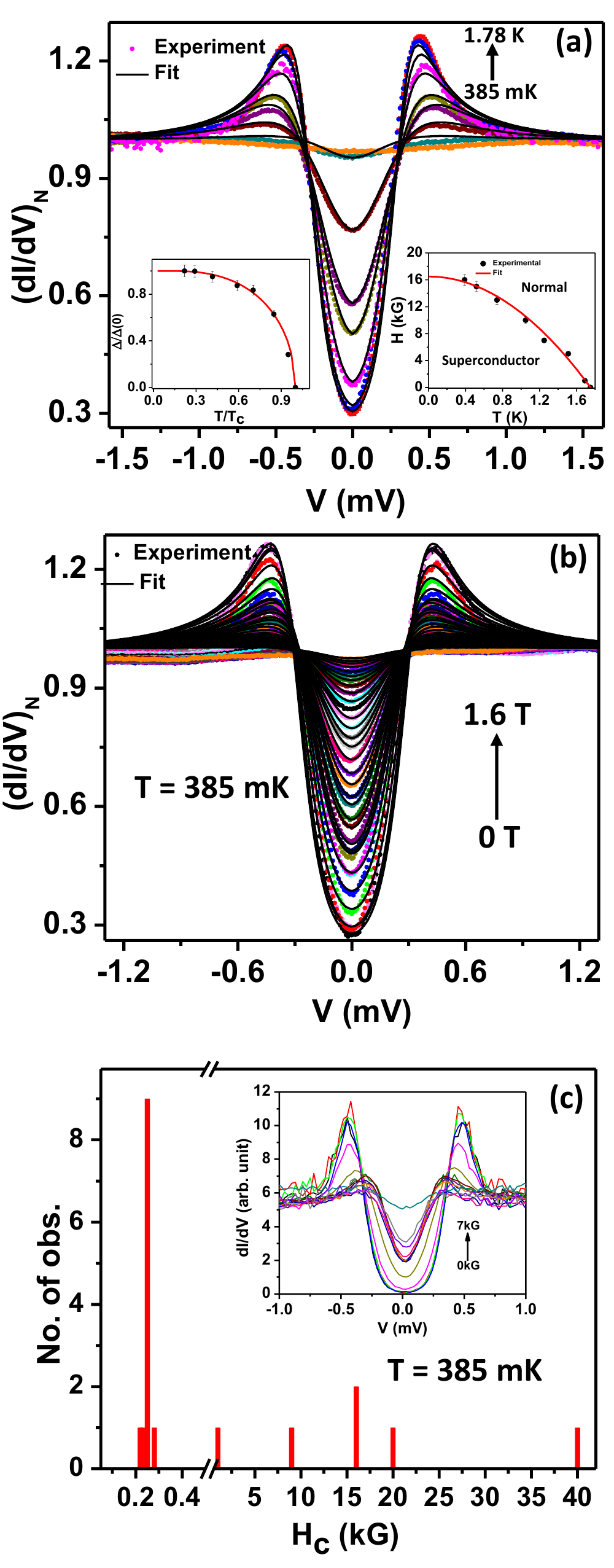}
	\caption{(a) Normalized STS data with varying temperature over a range from 385 mK to 1.78 K. The colored lines show experimental data points and the black lines show theoretical fits within BCS theory. inset (a) left: Temperature evolution of $\Delta$. The dots are values extracted from the theoretical fits and the solid line shows the temperature dependence as per BCS theory. inset (a) right: The H-T phase diagram extracted from the temperature dependent STS measurements at different magnetic fields. The solid line is a plot of the phase line expected empirically for conventional superconductors. (b) Normalized STS data with varying magnetic field up to 1.6 T. The colored lines show experimental data points and the black lines show theoretical fits within BCS theory. (c) Statistics of $H_c$ values recorded. $inset:$ Magnetic field dependence of the spectra with $H_c \sim 0.7$ T.}
	\label{Figure 4}
\end{figure}

In order to further investigate the superconductivity in PdTe$_2$, we carried out detailed temperature ($T$) dependence of the $dI/dV$ spectra. Figure 3(a) shows that with increasing temperature, the spectral features undergo broadening as expected. The position of the coherence peaks do not show noticeable change with increasing $T$ at low temperatures indicating the gap ($\Delta$) to be nearly constant at low temperature. As shown by the solid line in Figure 3(a), $\Delta$ decreases smoothly to zero following $\Delta(T)\sim \left(1-\frac{T}{T_c}\right)^{1/2}$  near $T_c \sim$ 1.78 K,\cite{CSyadav} as expected for a conventional BCS superconductor\cite{BCS}. The critical temperature thus measured is consistent with the value measured by temperature dependent heat capacity measurement on the same crystal ($inset$ of Figure 1(a)).


We have also investigated the behavior of the superconducting phase as a function of increasing magnetic field ($H$) (Figure 3(b). The magnetic field was applied along the $c$-axis of the crystal. The coherence peaks closed smoothly with increasing $H$ and all the spectral features disappeared at a high critical field $H_c$ of 1.6 Tesla at 385 mK (see Figure 3(b)). This corresponds to a coherence length of 14 nm. We have carried out temperature dependence of the spectra for different $H$ and constructed the $H-T$ phase diagram as presented in Figure 3(a) inset. The phase diagram shows a concave curvature with no visible tail near $T_c$ and matches well with the empirically expected behavior for a conventional superconductor. The solid line in Figure 3(a) inset represents the empirically expected phase line for a conventional superconductor where $H_c = H_0\left[1-\left(\frac{T}{T_c}\right)^2\right]$. The remarkable match of the experimental data with the theoretical line provides further support for the conventional nature of superconductivity in PdTe$_2$. This is surprising provided the existence of a topologically non-trivial band taking part in transport. The emergence of such a conventional superconducting phase might be due to electron pairing taking place in another band. This possibility cannot be ruled out since our dHvA oscillation measurements with $H||c$-axis, show that there are other bands available which could be trivial and could lead to superconductivity. Thus, from our measurements it is clear that despite the presence of at least one band with topologically non-trivial character, PdTe$2$ displays conventional superconductivity.

The critical field showed inhomogeneity from point to point on the cleaved PdTe$_2$ surface and it was possible to measure superconductivity up to 4 Tesla at certain points. This inhomogeneity could be due to a distribution of the local coherence length originating from the presence of randomly distributed impurities/defects. In Figure 3(c) we show a statistic on the observed value of the critical field. It is seen that while there is inhomogeneity, majority of the times, the critical field is found to be around 250 Gauss. This is in agreement with the bulk critical field measured by our magnetic field dependent heat capacity measurements and slightly higher than the reported value of bulk critical field by another group\cite{Leng}. No magnetic vortices were found in the conductance images recorded in presence of magnetic field. This is consistent with the previous claim of type I bulk superconductivity in PdTe$_2$\cite{Leng}. 


In conclusion, based on our detailed temperature and magnetic field dependent scanning tunneling microscopy and spectroscopy experiments, we have shown that contrary to the speculation of a topological superconducting phase, the superconducting phase emerging on the type II Dirac semimetal PdTe$_2$ is conventional in nature. The understanding of the mechanism for the emergence of unexpectedly conventional nature of the superconducting phase on PdTe$_2$ through detailed theoretical investigation could be a major step forward towards the understanding of superconductivity emerging in topologically non-trivial systems under different environments. 

We acknowledge the support of Mr. Avtar Singh, Ms. Astha Vasdev, Mr. Ritesh Kumar and Mr. Anzar Ali during various stages of this work. We acknowledge SEM and X-ray central facility at IISER Mohali. GS would like to acknowledge partial financial support from the research grant of (a) Swarnajayanti fellowship awarded by the Department of Science and Technology (DST), Govt. of India under the grant number DST/SJF/PSA-01/2015-16, and (b) the research grant from DST-Nanomission under the grant number SR/NM/NS-1249/2013.

\end{document}